\documentclass[aps,twocolumn,showpacs,amsmath,amssymb]{revtex4}
\usepackage{dcolumn}
\usepackage{bm}
\usepackage{color}
\usepackage{latexsym}
\usepackage{amssymb}
\usepackage{graphicx}
\usepackage{ulem}

\begin{document}
\title{Surprising coincidence of the apparent 2D metal-to-insulator transition data with Fermi-gas based model results}
\author{M. V. Cheremisin}

\affiliation{A.F.Ioffe Physical-Technical Institute, 194021
St.Petersburg, Russia}

\begin{abstract}
The melting condition for two-dimensional Wigner solid ( P.M. Platzman, H.Fukuyama, 1974) is shown to contain an error of a factor of $\pi$.
The analysis of experimental data for apparent 2D metal-to-insulator transition shows that the Wigner solidification
(B.Tanatar, D.M.Ceperley, 1989) has been never achieved. Within routine Fermi gas model both the metallic and insulating behavior of different
2D system for actual range of carrier densities and temperatures is explained.
\end{abstract}


\maketitle
Recently, much interest has been focused on the
anomalous transport behavior of a wide variety of low density
two-dimensional (2D) systems. It has
been found that below some critical density, cooling
causes an increase in resistivity, whereas in the opposite,
high-density case, the resistivity decreases. The apparent metal to insulator transition was observed in n-Si MOSFET \cite{Kravchenko95,Pudalov99},
p-GaAs\cite{Hanein98,Huang06,Huang11,Gao04},  n-GaAs\cite{Lilly03},  n-SiGe\cite{Lai05,Lai07} and
p-SiGe \cite{Coleridge97,Senz99} 2D systems.

\section{Wigner solidification vs experiment}
Let us provide the rigorous analysis of the experimental data regarding Fermi gas vs Wigner solid transition.
According to Ref.\cite{Platzman74}, the melting diagram of 2D Wigner solid obeys the condition
$\Gamma=E_{ee}/\langle K \rangle$, where $\Gamma$ is the coefficient assumed to be a constant at the phase
transition, $E_{ee}=\frac{e^{2}}{\epsilon}\sqrt{\pi N}$ is the Coulomb energy associated
to neighboring pair of electrons, $N$ is the 2D density. Within conventional Fermi gas model, ${\langle K \rangle}=kT\frac{F_{1}(1/\xi)}{F_{0}(1/\xi)}$ is
the average kinetic energy of single electron, where $F_{n}(z)$ is the Fermi integral of the order of $n$, $\xi =kT/\mu$ the
dimensionless temperature. Note that the average kinetic energy ${\langle K \rangle}$ coincides
with the thermal energy $kT$ for classical Boltzmann carriers $|\xi| \gg 1$. In contrast, ${\langle K \rangle}=\mu/2$ for
degenerate electrons $\xi \ll 1$. In general, the solidification of strongly degenerated electrons is believed to
occur at certain value of the Coulomb to Fermi energy ratio $r_{s}=E_{ee}/\mu$. We therefore conclude that $r_{s}=\Gamma/2$.
In Refs.\cite{Platzman74,Ando82}, this ratio has been erroneously defined as $r_{s}=\Gamma\frac{\sqrt{\pi}}{2}$, thus provides wrong
estimate for Wigner crystal solidification. For low-disorder 2D system Wigner solid was claimed\cite{Tanatar89} to exist when $r_{s}=37 \pm 5$.
\begin{figure}[tbp]
\begin{center}
\includegraphics[scale=0.6]{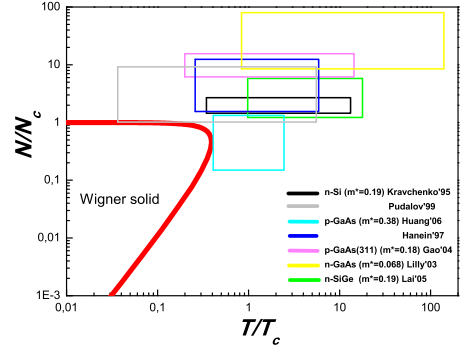}
\caption[]{\label{Fig1} The diagram of Fermi gas to Wigner solid transition \cite{Platzman74} according to
Eq.(\ref{Wigner_solid}) at $r_{s}=\frac{\Gamma}{2}=42$\cite{Tanatar89}. The color rectangular figures correspond to density and temperature range
of apparent metal-to insulator transition in Si-MOSFET\cite{Kravchenko95,Pudalov99};
p-GaAs\cite{Huang11,Gao04,Hanein98}; n-GaAs\cite{Lilly03} and n-SiGe\cite{Lai05,Lai07} 2D systems, modified
wit respect to dimensional density $N_{c}$ and temperature $T_{c}$ depicted in Table \ref{tab:table1}.}
\end{center}
\end{figure}

Following Ref.\cite{Platzman74}, the phase transition can be parameterized as it follows:
\begin{equation}
\frac{T}{T_{c}}=\frac{F_{0}^{3}(1/\xi)}{2F_{1}^{2}(1/\xi)}, \qquad \frac{N}{N_{c}}=\frac{F_{0}^{4}(1/\xi)}{4F_{1}^{2}(1/\xi)}.
\label{Wigner_solid}
\end{equation}
Here, the dimensional temperature $T_{c}=4\text{Ry}\frac{g_{v}}{\Gamma^{2}}$ and
2D density $N_{c}=\frac{4}{\pi a_{B}^{2}}\frac{g^{2}_{v}}{\Gamma^{2}}$ contain the valley splitting factor $g_{v}$. Then
$a_{B}=\frac{\epsilon\hbar^{2}}{me^{2}}$ and $\text{Ry}=\frac{me^{4}}{2\epsilon^{2}\hbar^{2}}$ is the effective Borh radius and Rydberg
energy respectively, $m$ is the effective mass. Note, for certain value of $r_{s}$ the correct values $T_{c},N_{c} \sim \Gamma^{-2}$
are lower by a factor of $\pi$ with respect to those predicted in Refs.\cite{Platzman74,Ando82}.
For actual 2D systems the values $T_{c},N_{c}$ are generalized in Table \ref{tab:table1}. In Fig.\ref{Fig1} we plot
the melting curve\cite{Platzman74} specified by Eq.(\ref{Wigner_solid}) and, moreover, the observed range of 2D densities and
temperatures attributed to apparent metal-insulator transition\cite{Kravchenko95,Pudalov99,Hanein98,Huang06,Huang11,Gao04,Lilly03,Lai05,Lai07}.
Evidently, Wigner solidification regime remains unaffected. Hence, we suggest the typical 2D systems can be described within routine Fermi gas model.
\begin{table}
\caption{\label{tab:table1} The density $N_{c}$ and temperature $T_{c}$ of Wigner phase transition at
$r_{s}=42$ for different 2D systems.}
\begin{tabular}{cccccccc}
2D system &$\varepsilon$&$m/m_{0}$&$g_{v}$&$N_{c}\cdot 10^{10}$ cm$^{-2}$&$T_{c}$,K&experiment\\
\hline   Si-MOSFET &7.7 &0.19 &2  &4.9   &0.6   & \cite{Kravchenko95},\cite{Pudalov99}   \\
         p-GaAs &13 &0.38  &1  &0.6   &0.2   & \cite{Hanein98},\cite{Huang11}   \\
         n-GaAs &13 &0.068 &1  &0.02  &0.04  & \cite{Lilly03} \\
         n-SiGe &11.7 &0.19 &2 &2.12  &0.25  & \cite{Lai05} \\
\end{tabular}
\end{table}

\section{Model of apparent metal-to insulator transition}
\begin{figure}[tbp]
\begin{center}
\includegraphics[scale=0.5]{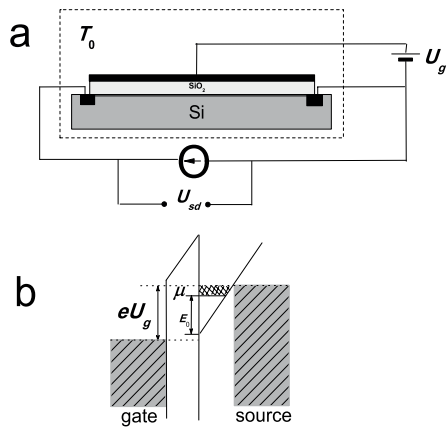}
\caption[]{\label{Fig2} a) The measurements setup; b) The band diagram of the of silicon metal-oxide-semiconductor FET system.}
\end{center}
\end{figure}

Let us first provide the arguments in favor of present use of Gibbs statistics. Since the experimental
observations concern the gate-based 2D system, we represent in Fig.\ref{Fig2} the sketch of the galvanic scheme and, moreover,
the related band diagram of typical Si-MOSFET structure. The latter is assumed to operate in strong inversion regime. The applied gate voltage
results in shift of Fermi level of 2D system with respect to that of the metal gate. The carrier density is given by the
number of occupied states below Fermi level counted from the bottom of the lowest subband of triangular quantum well. In thermodynamic
equilibrium the applied gate voltage is equal to electrostatic potential of the gate to channel capacitor, the chemical potential of
2D system $\mu/e$ and, finally, the flat-band potential $U_{t}=\Delta W/e$. Here, $\Delta W$ is the constant difference of the
work functions related to bulk silicon and metal gate respectively. The gate voltage $U_{g}$, shifted with respect to flat-band potential $U_{t}$ yields
\begin{equation}
U_{g}=\mathcal{Q}/C_{g}+\mu/e,
\label{diagram_condition}
\end{equation}
where $\mathcal{Q}=eN$ and $C_{g}=\epsilon/d$ is the charge density and geometrical capacity of the MOS capacitor respectively.
Then, $\epsilon$ and $d$ are the permittivity and thickness of the dielectric layer respectively. According to Eq.(\ref{diagram_condition}),
the increase(decrease) of the gate voltage results in change of the chemical potential and then, $indirectly$ in varying of 2D carrier density itself.
For strongly degenerated electrons the second term in Eq.(\ref{diagram_condition}) can be ignored providing the apparent evidence of 2D
density changed by the gate voltage. In contrast, for dilute 2D systems both terms in the right side of Eq.(\ref{diagram_condition})
have to be accounted. We therefore conclude that the applied gate voltage maintains the chemical potential of 2D system to be constant.
Consequently, the use of Gibbs statistics for 2D system is justified.

In Ref.\cite{Cheremisin05} low-T transport in dilute 2D systems has been analyzed
taking into account both the carrier degeneracy and
so-called thermal correction\cite{Kirby73} owing to Peltier and Seebeck
thermoelectric effects combined. For standard ohmic measurements setup,
the small applied current causes
heating(cooling) at the first(second) sample contact due to the
Peltier effect. Under adiabatic conditions the temperature
gradient is linear in current, the contact temperatures are
different. The measured voltage consists of the ohmic term and, moreover,
includes Peltier effect-induced
thermoemf which is linear in current. According to Ref.\cite{Cheremisin05},
the total measured resistivity yields
\begin{figure}[tbp]
\begin{center}
\includegraphics[scale=0.5]{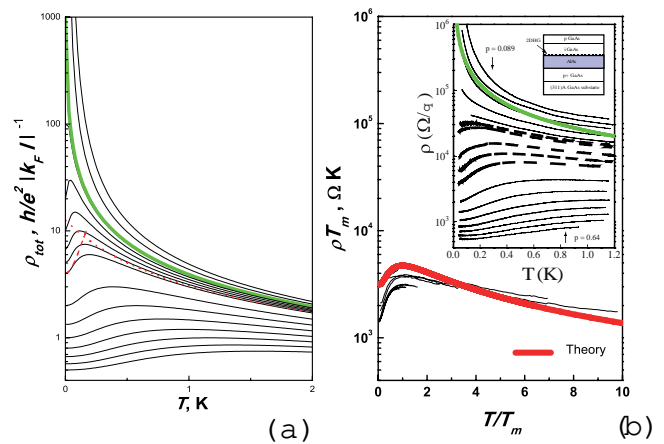}
\caption[]{\label{Fig3} a) Temperature dependence of 2DEG resistivity, given by Eq.(\ref{resistivity})
for $T_{F}[\text{K}] 2 - 0.25$ (step 0.25), 0.2-0.05( step 0.05), 0(green line),-0.1,-0.2 at fixed value of the
disorder strength: $k_{F}l=1$ at $T_{F}=1$K.
Dashed(doted) red line depicts the asymptote specified by Eq.(\ref{R assymptotes}) for degenerated $\xi \ll 1$ and non-degenerated $\xi >1$
gas respectively at fixed $T_{F}=0.5\text{K}$. b) Inset: experimental data\cite{Hanein98} for p-GaAs
system for carrier density $p=$0.089, 0.094, 0.099, 0.109, 0.119, 0.125, 0.13, 0.15, 0.17, 0.19, 0.25, 0.32, 0.38, 0.45, 0.51, 0.57
and 0.64 $\times 10^{11}$ cm $^{-2}$. The result of calculation for resistivity at $\mu=0$ is represented by green line.
Main panel: the result of scaling for curves(see insert), whose shapes demonstrate the
resistivity maxima for certain temperature within the range $0.12<T_{m}<1.64$K. Bold red line corresponds to universal dependence specified in text. }
\end{center}
\end{figure}
\begin{equation}
\rho_{tot}=\rho \left( 1+\alpha ^{2}/L\right),
\label{resistivity}
\end{equation}
where $\rho=\frac{m}{Ne^{2}\tau}$ is the ohmic resistivity, $\alpha$ is 2D
thermopower, $L=\frac{\pi ^{2}k^{2}}{3e^{2}}$ is the Lorentz number. For simplicity, we further
assume that the momentum relaxation time $\tau$ is energy-independent. Using Gibbs statistics
and parabolic energy spectrum for 2D carriers we obtain $N=N_{0}\xi F_{0}(1/\xi )$, where $N_{0}=\frac{g_{v} m \mu }{\pi \hbar ^{2}}$
is the density of strongly degenerated electrons. Then, within Boltzman equation
formalism the 2DEG thermopower (for 3D case, see Pisarenko, 1940) yields $\alpha =-\frac{k}{e}
\left[ \frac{2F_{\mathrm{1}}(1/\xi )}{F_{0}(1/\xi )}-\frac{1}{\xi }
\right] $.

As it was demonstrated in Ref.\cite{Cheremisin05}, Eq.(\ref{resistivity}) provides the faithful
sketch for transport behavior of dilute 2D systems. Both the theory results and experimental data\cite{Hanein98}
are shown in Fig.\ref{Fig3}. At first, we are interested in the case of Fermi level lying
above the bottom of the conducting band $\xi\geq 0$, which corresponds to portion of data below green line separatrix. One can
distinguish the puzzling temperature behavior of the resistivity as $\frac{\partial\rho}{\partial T}>0$ for degenerated carriers
$\xi \ll 1$ and $\frac{\partial\rho}{\partial T}<0$ for high-T case $\xi \geq 1$. The resistivity data ( see, for example, the curve
at $T_{F}=0.25K$ in Fig.\ref{Fig3}) exhibits the maximum $\rho_{m} = 1.5 \rho_{0}$ at $T_{m}=0.78T_{F}$. These values are close to
those observed experimentally. The subsequent asymptotes for total resistivity specified by Eq.(\ref{resistivity}) can be written as
\begin{eqnarray}
\xi \ll 1, \qquad \rho_{tot}=\rho_{0}(1+\pi ^{2}\xi ^{2}/3), \\
\label{R assymptotes}
\xi \geq 1, \qquad \rho_{tot}=\rho_{0}\frac{ 1+\alpha _{s}^{2}/L}{1/2+\iota+\xi\ln 2}.\\
\nonumber
\end{eqnarray}
Here, $\rho _{\mathrm{0}}\mathbf{=}\frac{h}{e^{2}}
|k_{F}l|^{-1}$ is the ohmic resistivity at $T\rightarrow
0 $,  $k_{F}=\sqrt{2m |\mu| }/\hbar $ is the Fermi vector, $l_{p}=$ $\hbar k_{F}\tau /m$ is the mean free path.
Then, $\alpha _{s}=-\frac{k}{e}\frac{\pi ^{2}}{6\ln 2}$ is the
thermopower for non-generated carriers $\xi \gg 1$, $\iota=\frac{\alpha _{s}^{2}/L-1}{\alpha _{s}^{2}/L+1}$ is a correction.

\begin{figure}[tbp]
\begin{center}
\includegraphics[scale=0.55]{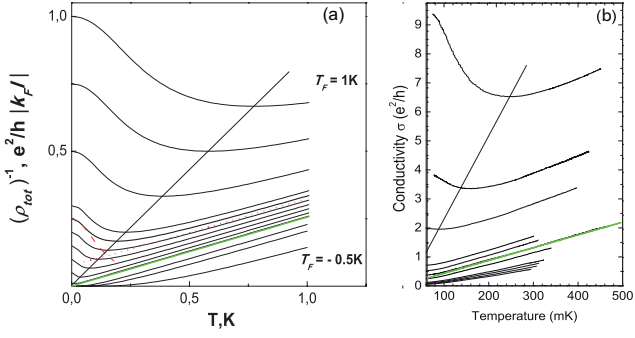}
\caption[]{\label{Fig4} a) The inverse resistivity $\rho_{tot}^{-1}$ vs temperature re-plotted from main
panel of Fig.\ref{Fig3}a. Straight line depicts the minima position.
b) Experimental data $\rho_{tot}^{-1}(T)$ for p-GaAs system\cite{Huang11} for 2D hole density $0.8, 1.2, 1.6, 1.8, 2.5, 3.0, 3.4,
3.8, 5.1, 6.1, 7.1 \times 10^{9}$ cm $^{-2}$. The green straight line depicts the expected dependence for $T_{F}=0$.}
\end{center}
\end{figure}

Let us examine in details the behavior of 2D resistivity shown in Fig.\ref{Fig3}. One may check whether the maxima positions
of the resistivity curves obey the predicted relationship $\rho_{m} \sim 1/T_{m}$. To confirm this, in Fig.\ref{Fig4} we plot
the inverse total resistivity $\rho_{tot}^{-1}$ vs temperature and, moreover, re-plot the experimental
data\cite{Huang11} shown in Fig.\ref{Fig3}. Surprisingly, both the theory and experiment follow the expected linear
dependence $\rho_{m}^{-1}\sim T_{m}$. Based on these findings, we suggest a simple scaling procedure which can be
applied to original experimental data. Indeed, for certain resistivity curve demonstrated a certain
maximum at $T_{m}$ one can find the product $\rho_{tot} T_{m}$
which is presumably universal function of ratio $T/T_{m}$. The data set\cite{Hanein98} scaled in this manner is represented
in Fig.\ref{Fig3}b. Remarkably, the original two-order of magnitude range data shrink to roughly unique curve.
In addition, we put on the same plot the result provided by theory. Indeed, for hole density $p=0.64 \times 10^{11}$ cm$^{-2}$
reported in Ref.\cite{Hanein98} the extrapolation $T\rightarrow 0$ yields the resistivity $\rho_{0}=530\Omega$. The respective carrier
mobility $\mu=1.85\times 10^{5}$cm$^{2}$/Vs corresponds to Dingle temperature $T_{D}=\frac{\hbar}{k\tau}=0.19$K. Consequently, in
Fig.\ref{Fig3},b we plot the dimensional dependence of the product $\rho_{tot}T_{m}=\frac{h}{e^{2}}\frac{T_{D}}{2\cdot 0.78}w(T/T_{m})$, where
$w(\xi)=(1+\alpha^{2}/L)(\xi F_{0}(1/\xi))^{-1}$ is the universal function. Data scaling result agrees with that provided by theory.

\begin{figure}[tbp]
\begin{center}
\includegraphics[scale=0.7]{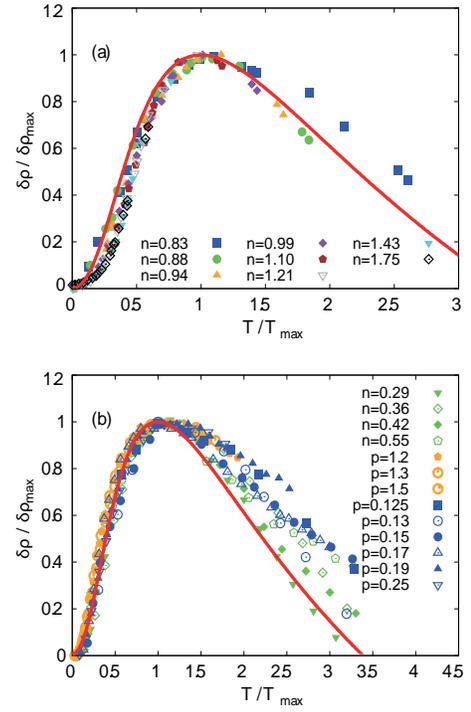}
\caption[]{\label{Fig5} Scaling anzatz of 2D resistivity data \cite{Radonjic12} for: a) Si-MOSFET\cite{Pudalov99}
b) p-GaAs/AlGaAs(black symbols) \cite{Hanein98}, n-GaAs/AlGaAs(green symbols)\cite{Lilly03},
p-GaAs(orange)\cite{Gao05}. The the red line represents the universal function $W(T/T_{m})$ extracted from Eq.(\ref{resistivity}). }
\end{center}
\end{figure}

We now verify validity of our approach for different 2D systems based on scaling procedure made of use
in Ref.\cite{Radonjic12}. The experimental data argued to exhibit a certain universality as
function of reduced temperature $T/T_{m}$ and using the dimensionless form $\frac{\rho_{tot}-\rho_{0}}{\rho_{m}-\rho_{0}}$.
Note that in our notations the latter ratio is equal to $2(w-1)$, thus support the universality.
Fig.\ref{Fig5} represent the scaling analysis\cite{Radonjic12} for different 2D systems. Additionally, in Fig.\ref{Fig5} we put
the result of calculations within our theory. The coincidence between theory and experiment is rather impressive.

Further progress in verifying whether our model is correct concerns the high-T dependence of the resistivity.
Inverting Eq.(\ref{R assymptotes}) we obtain the linear dependence of inverse total resistivity $\rho_{tot}^{-1}(T)$ as
\begin{figure}[tbp]
\begin{center}
\includegraphics[scale=0.5]{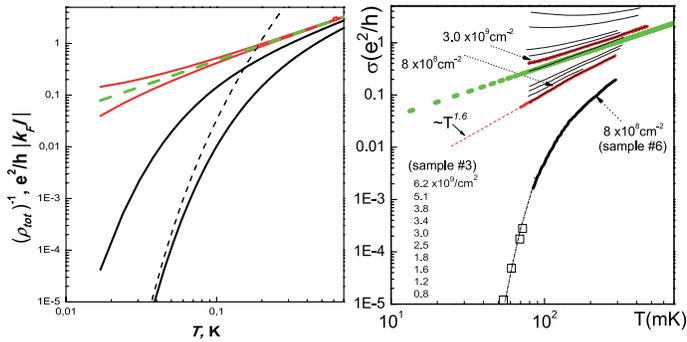}
\caption[]{\label{Fig6} a) The temperature dependence of inverse total resistivity
$\rho_{tot}^{-1}$ for different values of the Fermi temperature $T_{F}[K]=0.01; -0.01;-0.1;-0.3$ and fixed Dingle temperature $T_{D}=0.11$K.
The dotted line corresponds to $T_{F}=0$. The dashed line represents the asymptote specified by Eq.(\ref{resistivity_activated}). b) Re-plot of experimental
data\cite{Huang06,Huang11} shown in Fig.\ref{Fig4} b, for p-GaAs sample($T_{D}=0.11$K, see Table \ref{tab:table1}) for hole density $ 6.2, 5.1, 3.8, 3.4, 3.0, 2.5, 1.8, 1.6, 1.2, 0.8 \times 10^{9}$ cm $^{-2}$.
The dotted green line corresponds to zero Fermi energy. The bold black line depicts the insulating behavior of
high-disordered sample studied in Ref.\cite{Huang06}.}
\end{center}
\end{figure}
\begin{eqnarray}
\rho_{tot}^{-1}=\mathcal{A}+\mathcal{B} \cdot T\\
\label{invers_resistivity}
\mathcal{A}=\frac{\sigma_{0}(1/2+\iota)}{1+\alpha^{2}_{s}/L},\qquad \mathcal{B}=\frac{e^{2}}{h} \frac{1}{T_{D}}\frac{2g_{v}\ln2}{1+\alpha^{2}_{s}/L}.
\nonumber
\end{eqnarray}
Here, $\sigma_{0}=1/\rho_{0}$ is 2D conductivity at zero temperature.
High-T linear behavior of inverse total resistivity is clearly seen in experiment (see Fig.\ref{Fig4},b) and, therefore, available
for qualitative analysis with respect to theory. The estimates for temperature coefficient $\mathcal{B}$ are summarized
in Table \ref{tab:table0} being in agreement with theory predictions. Note that zero Fermi energy case $\mu=0$ (see green lines in Fig.\ref{Fig4})
is of special interest since the inverse total resistivity $ \rho_{tot}^{-1}=\mathcal{B} \cdot T$ vanishes at $T\rightarrow 0$.
Simultaneously, for $\mu=0$ the resistivity is expected to hyperbolic increase at $T\rightarrow 0$ clearly seen in
Fig.\ref{Fig3}b, insert for p-GaAs system\cite{Hanein98}.

Finally, we consider the most intriguing case of Fermi level laying below the bottom of the conducting band, i.e. when $\xi<0$.
This regime corresponds to portion of data below green line seperatix in Fig.\ref{Fig4}. For actual strong insulating case
$|\xi|\ll 1$ the 2D density is exponentially small $\sim |\xi|\exp(-1/|\xi|)$, while the the thermopower behavior yields
the Boltzman form $\alpha\sim \frac{k}{e}\frac{1}{\xi}$. Finally, we obtain the asymptote for 2D resistivity as it follows
\begin{equation}
\xi<0, |\xi| \ll 1, \qquad \rho_{tot}=\rho_{0}\frac{3}{\pi^{2}} |\xi|^{-3}\exp(1/|\xi|)\\
\label{resistivity_activated}
\end{equation}
In order to compare our results with experimental data, in Fig.\ref{Fig6}a we present the log-log version of Fig.\ref{Fig4}a.
As expected, the curves demonstrate the progressive change from high-T linear trend to low-T activated behavior. As expected, low-T
activated behavior is described by inverted Eq.(\ref{resistivity_activated}). Surprisingly, the typical experimental results\cite{Huang06} shown
in Fig.\ref{Fig6}b demonstrate similar behavior. Note that usually there exist an overall failure\cite{Huang06} to fit low-T data(see in Fig.\ref{Fig6}b) within
variable range hopping formalism $\rho_{tot}^{-1} \sim e^{(-T^{*}/T)^{\nu}}$, where $\nu=1/3$ and $\nu=1/2$ corresponds to Mott\cite{Mott68}
and Efros-Shklovskii\cite{Shklovskii84} predictions respectively.
\begin{table}
\caption{\label{tab:table0} Temperature coefficient $\mathcal{B}$}
\begin{tabular}{cccccccc}
2D system    &$\mu$, m$^{2}$/Vs &$T_{D}$,K &$\mathcal{B}, \frac{e^{2}}{h} K^{-1}$(exp/th)&ref\\
\hline   Si-MOSFET &19 &0.38                     &2.7/0.4        & \cite{Lai07}   \\
         p-GaAs &28 &0.13                     &4.0/3.0        & \cite{Gao03}   \\
         p-GaAs &32 &0.11                     &4.7/4.5        &  \cite{Huang06},\cite{Huang11}  \\
\end{tabular}
\end{table}

In conclusion, we demonstrate that Wigner solidification has been never achieved in experiments dealt
with apparent metal to insulator transition. The observed anomalies of 2D transport behavior is explained within conventional
Fermi gas formalism invoking the important correction to measured resistivity caused by Peltier and Seebeck
effects combined. We represent the experimental evidence confirming the solidity and universality of the above model.

\end{document}